\documentclass[final,3p,times,twocolumn]{elsarticle}

 \biboptions{comma,sort&compress}
\usepackage{here}
 \usepackage{graphicx}
  \usepackage{epsfig}

\usepackage{amsmath}

\def\nin{\noindent}
\def\beq{\begin{equation}}
\def\eeq{\end{equation}}
\def\bea{\begin{eqnarray}}
\def\eea{\end{eqnarray}}

\journal{Nuc. Phys. (Proc. Suppl.)}

\begin{document}

\begin{frontmatter}



\title{Electromagnetic corrections in hadronic $\tau$ decay}

\author[label1]{F.V. Flores-Ba\'ez\corref{cor1}}
\address[label1]{Instituto de F\'isica Corpuscular, IFIC.
\\ Universitat de Val\`encia-CSIC\\
Apartat de Correus 22085, 
E-46071, Val\`encia, Spain.}
\cortext[cor1]{Speaker}
\ead{fvflores@ific.uv.es}



\begin{abstract}
\noindent
It is well known that in every experiment involving charged particles there are individual photons  emitted with  small energy $\omega$ and those with $\omega\leq \omega_{0}$ will not be detected but are present in the data. The typical procedure to estimate this effect consists in taking the leading logarithm from the soft photon approximation and to make a simulation through a Monte Carlo Algorithm (PHOTOS). The aim of this work is to compute the model independent  QED correction to $\tau\to K\pi\nu$ decay and to estimate its effect in the decay width in the particular process $\tau^{+}\to K^{0}\pi^{+}\nu$.

\end{abstract}


\end{frontmatter}


\section{Introduction}
\nin
Hadronic $\tau$ decays are an ideal laboratory to obtain information about the fundamental parameters within the Standard Model and also some properties of QCD at low energies \cite{sm}\cite{qcd}.
In particular the $\tau\to K\pi \nu$ decay has been studied in the past by ALEPH \cite{aleph} and OPAL \cite{opal} and recently at B factories \cite{babar}\cite{belle} where the high statistic measurements provide excellent information about the structure of the spectral functions, parameters of the intermediate states and the total hadronic spectral function. Also it is possible to determine the product $|V_{us}|F_{+}^{K^{0}\pi^{+}}(0)$ from this decay although the best determination comes from semileptonic kaon decays \cite{antonelli}. 

The nice feature of this sort of process is that the decay amplitude can be factorized into a pure leptonic part and a hadronic spectral function \cite{mirkes} in such a way that the differential decay distribution reads
\begin{align}
\frac{d\Gamma_{K\pi}}{d\sqrt{s}}&=\frac{G_{F}^{2}\left(|V_{us}|F_{+}^{K\pi}(0)\right)^{2}m^{3}_{\tau}}{32\pi^{3}s}S_{EW}\left(1-\frac{s}{m^{2}_{\tau}}\right)^{2}\nonumber\\
&\times\left[\left(1+\frac{2s}{m^{2}_{\tau}}\right)q^{3}_{K\pi}|\tilde{F}_{+}^{K\pi}(s)|^{2}
+\frac{3\Delta_{K\pi}^{2}q_{K\pi}|\tilde{F}_{0}^{K\pi}(s)|^{2}}{4s}\right]\ ,\nonumber\\
\end{align}
where a sum over the two posibles decays $\tau^{+}\to K^{+}\pi^{0}\nu$ and $\tau^{+}\to K^{0}\pi^{+}\nu$ has been done, isospin symmetry has been assumed and the reduced vector and scalar form factors have been normalized to one at the origin
\begin{align}
\tilde{F}_{+}^{K\pi}(s)&= \frac{F_{+}^{K\pi}(s)}{F_{+}^{K\pi}(0)}\ ,\qquad \tilde{F}_{0}^{K\pi}(s)= \frac{F_{0}^{K\pi}(s)}{F_{+}^{K\pi}(0)}\ .
\end{align}
In this expresion $\Delta_{K\pi}\equiv m^{2}_{K}-m^{2}_{\pi}$  and the kaon momentum in the rest frame of the hadronic system is
\begin{align}
q_{K\pi}&=\frac{1}{2\sqrt{s}}\sqrt{(s-(m_{K}+m_{\pi})^{2})(s-(m_{K}-m_{\pi})^{2})}\nonumber\\
&\times \theta(s-(m_{K}+m_{\pi})^{2})\ .
\end{align}
Theoretical descriptions \cite{jamin} to the vector and scalar form factors have been done in the Chiral Theory with Resonances (R$\chi$T) framework \cite{rchpt} providing a succesful representation of the data. It is worth to mention that  Eq. (1) includes the short distance correction $S_{EW}$ \cite{erler} but the long distance correction is missed.\\
It is well known that in all decays  with a charged particle an emission of photons is always present altering  the dynamics of the decay. A complete theoretical treatment including the next-to-leading order electromagnetic effects in hadronic $\tau$ decays is required  whenever an impressive experimental precision has been achieved.\\
In the past a simple  effective interaction approach  was  used for a treatment of lepton-hadron interactions, however once  the electromagnetic corrections are computed an ultra-violet (UV) singularity appears \cite{ginsberg}.\\
In the Chiral Perturbation Theory (ChPT) framework \cite{chpt} it is possible to describe hadron-lepton interactions at low energy including also real and virtual photons and leptons \cite{chptconfotones}\cite{chptleptons} without the UV problem, as can be seen in  the nice work done in  $K_{l3}$ \cite{kl3}.\\
In addition in a radiative process a photon with energy $\omega\leq \omega_{0}$ is not detected but affects the measurements\footnote{ A photon with energy $\omega\leq\omega_{0}$ is called soft provided that $\omega_{0}$ is smaller than typical energies present in the process.}. The approximative next-to-leading  order algorithms \cite{photos} are used to simulate the correction due to  soft photons where  the virtual  corrections (one loop) are  reconstructed numerically up to the leading logarithms from the real photon corrections. Also an improved algorithm \cite{klor} can be applied if a phenomenological model is used to describe the behaviour of the invariant amplitude.\\
This work  presents the Model Independent (MI) electromagnetic  corrections due to photons that are not detected (virtual  and soft-real photon ). In Section 2 is described  briefly  the basis of the hadronic tau decay and  general effects of electromagnetic corrections. In Section 3 the MI electromagnetic corrections are computed. In Section 4 these MI corrections to the $\tau^{+}\to K^{0}\pi^{+}\nu$ decay width are presented.

\section{$\tau$ decay and the Electromagnetic Interaction}

The invariant amplitude  for the $\tau$ decay into the hadronic state $K\pi$  can be expressed \cite{mirkes} as follows
\begin{eqnarray}\label{eq:mglevel}
\mathcal{M}_{\tau}&=& CG_{F}V_{us}l_{\mu}h^{\mu}\ ,
\end{eqnarray}
where the leptonic current is given by 
\begin{eqnarray}
l_{\mu}&=&\bar{\nu}_{\tau}\gamma_{\mu}\gamma^{7}u_{\tau}\ ,
\end{eqnarray}
and the hadronic part $h^{\mu}$ that contains the form factors $F_{+}(s)$ and $ F_{-}(s)$, is written as follows
\begin{eqnarray}
h^{\mu}=F_{+}(s)(p_{+}-p_{0})^{\mu}+F_{-}(s)(p_{+}+p_{0})^{\mu}\ ,
\end{eqnarray}
where $ p_{+}(p_{0})$ is the 4-momenta of the charged (neutral) scalar meson, $s=(p_{+}+p_{0})^{2}$ and the Clebsch-Gordon coefficient $C=(-\frac{\sqrt{2}}{2},\frac{1}{2})$ depends on the hadronic final state ($\pi^{+}K^{0}, \pi^{0}K^{+}$).

In the chiral limit  the tree level invariant amplitude reads
\begin{eqnarray}\label{eq:tlevel}
\mathcal{M}^{(0)}_{\tau}=CG_{F}V_{us}l_{\mu}(p_{+}-p_{0})^{\mu}\ .
\end{eqnarray}
Considering only the EM interactions, the total amplitude is found to be
\begin{align}
\mathcal{M}_{\tau}&=CG_{F}V_{us} l_{\mu}\mathcal{H}^{\mu}_{e.m.}\ ,\nonumber\\
\mathcal{H}^{\mu}_{e.m.}&=\left[F_{+}(u)(p_{+}-p_{0})^{\mu}+F_{-}(u)(p_{+}+p_{0})^{\mu} \right]\ ,\nonumber\\
F_{+}(u)&= 1 + f^{e.m.}_{+}(u)\ ,\nonumber\\
F_{-}(u)&= f^{e.m.}_{-}(u)\ ,
\end{align}
where $u=(P_{\tau}-p_{+})^{2}$ and $f^{e.m.}_{\pm}(u)$ contains the electromagnetic corrections. In the power counting of ChPT, the electromagnetic interaction produces a correction to the form factors that starts at $\mathcal{O}(e^{2}p^{0})$ \cite{kl3b}.
In general  $f^{e.m.}_{+}(u)\neq f^{e.m.}_{-}(u)$ and they have an UV singularity. To solve the UV problem there are at least two procedures:
\begin{itemize}
\item To apply ChPT with resonance (R$\chi$T) including virtual photons and leptons.
\item To choose a cut-off scheme in the 4-dimensional integration.
\end{itemize}

The first point is the most suitable approach, however the coupling constant of the  lagrangian are unknown from theory but can be determined by low-energy phenomenology. That exhaustive task is out of the scope of this work. We start with the amplitude at lowest order Eq. (\ref{eq:tlevel}) and following the approach of the ref.\cite{Yena} a Model Independent (MI)  electromagnetic correction is computed, which is free of UV singularities and it does not depend of details of the interactions.
\nin
\section{Model Independent  Radiative Corrections}
In this section the Model Independent one loop and real photon correction is computed following the techniques explained in ref.\cite{Yena}.
\subsection{Virtual photons}
Starting with the tree level amplitude  Eq. (\ref{eq:tlevel}) and using pointlike interaction between  photon and charged hadron, the MI one loop electromagnetic correction is obtained from the Feynman diagrams of Fig. (1) .\\
\begin{figure}[hbt]
\centerline{\includegraphics[width=.90\linewidth]{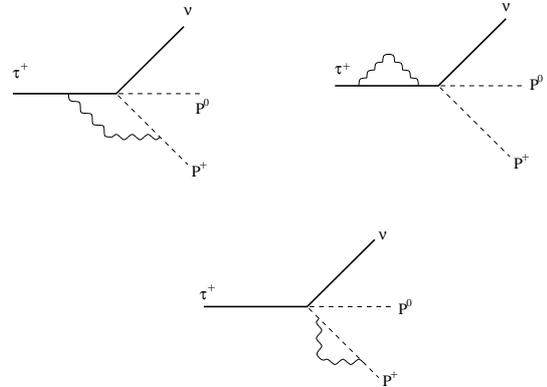}}
\caption{\scriptsize The so-called external diagrams, where a photon line conects external charged legs, $P^{+}(P^{0})$ denotes the charged(neutral) scalar meson.}
\label{fig1}
\end{figure}
Hence the total amplitude of the MI electromagnetic corrections takes the form
\begin{align}\label{eq:oneloop}
\mathcal{M}_{m.i.}^{v}&=\mathcal{M}^{(0)}_{\tau}\frac{\alpha}{4\pi}f^{v}_{\rm m.i.}\ ,\nonumber\\ f^{v}_{\rm m.i.}&= 2\ln\left[\frac{m_{+}m_{\tau}}{\lambda^{2}}\right]-\frac{m^{2}_{+}}{2u}\left(1-r_{t}\right)\ln[r_{t}]-1\nonumber\\ &   +\frac{m_{+}m_{\tau}}{u}\left[\frac{1}{x_{t}}-x_{t}\right]\ln[x_{t}]\nonumber\\&+2m^{2}_{+}y_{t}C0[m^{2}_{+},u,m^{2}_{\tau},\lambda^{2}]\ .
\end{align}
The notation is the same that ref. \cite{cirigliano}, $\lambda$ is the photon mass and the expression of $C0[m^{2}_{+},u,m^{2}_{\tau},\lambda^{2}]$ is obtained from the general form given in \cite{denner}.

\nin
\subsection{Real Soft Photon}
In order to eliminate the IR-divergences due to the vanishing of the photon mass, it is necessary to take into account the process with a single photon emission \cite{bloch}.\\
According to the Low Energy Theorems \cite{low}\cite{kroll} we define our MI real photon amplitude that is also gauge invariant as follows
\begin{equation}\label{eq:bk}
\mathcal{M}^{\gamma}_{\tau}=\mathcal{M}^{(0)}_{\tau} e \left[\frac{p^{+}\cdot \epsilon(k)}{p^{+}\cdot k}-\frac{P_{\tau}\cdot\epsilon(k)}{P_{\tau}\cdot k}\right]\ .
\end{equation}
The  differential width of the radiative process can be separate as follows
\begin{eqnarray}\label{eq:spb}
d\Gamma^{m.i.}_{\tau\to K\pi\nu\gamma}=d\Gamma^{m.i.}_{\tau\to K\pi\nu\gamma}|^{\omega_{0}}_{0}+d\Gamma^{m.i.}_{\tau\to K\pi\nu\gamma}|_{\omega_{0}}^{\omega_{max}}\ ,
\end{eqnarray}
where $\omega_{max}=[m^{2}_{\tau}-(m_{+}+m_{0})^{2}]/2m_{\tau}$. The first part is the focus of this work, the soft photon correction. The  MI differential decay width due to soft photon in the $\tau\to K\pi\nu_{\tau}\gamma$  process reads
\begin{eqnarray}
d\Gamma^{m.i.}_{\tau\to K\pi\nu\gamma}&=& \frac{|C|^{2}G^{2}_{F}\mathcal{D}(t,u)dsdu}{m^{3}_{\tau}(4\pi)^{3}}\left[\frac{\alpha}{\pi}g^{\gamma}_{soft}\right]\ ,
\end{eqnarray}
where the cinematic density is 
\begin{eqnarray}
\mathcal{D}(s,u)&=& 2u^{2}+\frac{m^{2}_{\tau}}{2}(m^{2}_{\tau}-s)+2m^{2}_{+}m^{2}_{0}\nonumber\\& &+2u(s-m^{2}_{\tau}-m^{2}_{+}-m^{2}_{0})\ .
\end{eqnarray}
The function $g^{\gamma}_{soft}$ includes the infrared term that has been extracted carefully \cite{coester} and reads\footnote{The dilogarithm function is $Li_{2}[x]=-\int_{0}^{x}dt\ln [1-t]/t$}
\begin{align}\label{eq:sp}
g^{\gamma}_{soft}&= 1+\ln\left[\frac{\lambda}{\omega_{0}}\right]\left[2+\frac{2x_{t}y_{t}\ln[x_{t}]}{\sqrt{r_{t}}[1-x^{2}_{t}]} \right]\nonumber\\&-\frac{y_{t}x_{t}}{2\sqrt{r_{t}}[1-x^{2}_{t}]}\left\{\ln[4]\ln[x^{2}_{t}]+\rm Li_{2}\left[1-x^{2}_{t}\right]\right.\nonumber\\& \left.
-\rm Li_{2}\left[\frac{x^{2}_{t}-1}{x^{2}_{t}}\right]+\ln[x^{2}_{t}]  \right\} -2\ln[2]\ .
\end{align}
The MI electromagnetic correction reads
\begin{eqnarray}\label{eq:miec}
\Delta^{e.m.}_{m.i.}(m_{\tau},m_{+},u,\omega_{0})&=&\frac{\alpha}{\pi}[g^{\gamma}_{soft}+\frac{f^{v}_{\rm m.i.}}{2}]\ ,
\end{eqnarray}
where the infrared singularity  is canceled as it should be. 

\nin
\section{Correction to the decay}
The differential decay width  without electromagnetic corrections can be written as follows
\begin{eqnarray}
\frac{d\Gamma_{ K\pi}}{dsdu}&=& \frac{G^{2}_{F}|V_{us}F^{K\pi}_{+}(0)|^{2}C^{2}}{(4\pi)^{3}m^{3}_{\tau}}\rho(s,u)\ ,
\end{eqnarray}
where the  density is
\begin{eqnarray}\label{eq:density}
\rho(s,u)&=& |\tilde{F}_{+}(s)|^{2}\mathcal{D}(s,u)+|\tilde{F}_{-}(s)|^{2}\mathcal{D}_{2}(s)\nonumber\\& & +2 \tilde{F}_{-}(s)\tilde{F}_{+}(s)\mathcal{D}_{3}(s,u)\ ,\nonumber\\
\mathcal{D}_{2}(s)&=&\frac{m^{2}_{\tau}}{2}(m^{2}_{\tau}-s)\ ,\nonumber\\
\mathcal{D}_{3}(s,u)&=&\frac{m^{2}_{\tau}}{2}[2m^{2}_{0}+m^{2}_{\tau}-s-2u]\ ,
\end{eqnarray}
In order to include the MI electromagnetic interactions we follow the approach explained in ref. \cite{kl3b}. Assuming that QCD and EM effects are turned on, the form factors are expanded according to the chiral properties  as follows
\begin{eqnarray}
F_{+}&=& 1+\mathcal{O}(p^{2})+ \mathcal{O}(e^{2}p^{0})_{m.d.}^{+} +\mathcal{O}(e^{2}p^{0})^{+}_{m.i.}\ ,\nonumber\\
F_{-}&=& \mathcal{O}(p^{2}) + \mathcal{O}(e^{2}p^{0})^{-}\ .
\end{eqnarray}
The electromagnetic effects in the form factors have been considered at the lowest order $\mathcal{O}(e^{2}p^{0})$, also in $F_{+}$ the MI electromagnetic correction, described in the previous section, has been separated from the model dependent correction. It is assumed that all the details of how to deal with the UV singularity and the structure behaviour of the photon-hadron interaction are left in the model dependent form factor.
Hence  the  differential decay width including the MI electromagnetic corrections is found to be
\begin{align}\label{eq:total}
\frac{d\Gamma_{ K\pi}}{dsdu}&=\frac{G^{2}_{F}|V_{us}F^{K\pi}_{+}(0)|^{2}C^{2}}{(4\pi)^{3}m^{3}_{\tau}}\left\{  |\tilde{F}_{+}(s)|^{2}\mathcal{D}(s,u)\right.\nonumber\\& \left.+ |\tilde{F}_{-}(s)|^{2}\mathcal{D}_{2}(s)+2 \tilde{F}_{-}(s)\tilde{F}_{+}(s)\mathcal{D}_{3}(s,u)\right.\nonumber\\& \left. +\mathcal{D}(s,u)\Delta^{e.m.}_{m.i.}(m_{\tau},m_{+},u,\omega_{0}) \right\}\ ,
\end{align}
where the integrated density is defined as follows
\begin{eqnarray}
I_{K\pi}&=& \int  \rho(s,u) ds du\ .
\end{eqnarray}
Integrating Eq.(\ref{eq:total}) allows us to know the size of the MI electromagnetic corrections,
\begin{align}
\Gamma_{K\pi}&=\frac{G^{2}_{F}|V_{us}F^{K\pi}_{+}(0)|^{2}C^{2}I_{K\pi}}{(4\pi)^{3}m^{3}_{\tau}}\left\{ 1+ \delta_{m.i.}^{e.m.} \right\}\ ,\nonumber\\
\delta_{m.i.}^{e.m.}&=\frac{1}{I_{K\pi}}\int \mathcal{D}(s,u)\Delta^{e.m.}_{m.i.}(m_{\tau},m_{+},u,\omega_{0})dsdu\ . \nonumber\\
\end{align}
Once $I_{K\pi}$ is computed, the size of these MI soft electromagnetic corrections can be known. Considering the decay $\tau^{+}\to K^{0}\pi^{+}\nu$, using  diferent values of $\omega_{o}$ and $I_{K^{0}\pi^{+}}=0.802356$ \cite{passemar}, the size of $\delta_{m.i.}^{e.m.}$ is shown in Table 1.

{\scriptsize
\begin{table}[hbt]
\setlength{\tabcolsep}{.5pc}
\caption{\scriptsize Size of the MI Electromagnetic soft corrections in $\tau^{+}\to K^{0}\pi^{+}\nu$. }
    {\small
\begin{tabular}{cccccc}
&\\
\hline
$ \omega_{0}$(MeV)& $10$ & $20$ & $30$& $40$&  $50$ \\
\hline
$\delta^{e.m.}_{m.i.}$ & -0.052 & -0.041& -0.035 &-0.031 &-0.027\\ \hline
\end{tabular}
}
\label{tab:param}
\end{table}
}
\nin
\section{Conclusions}
\nin
Summarizing the work done, the procedure described in ref. \cite{Yena} is used to estimate the MI electromagnetic corrections to the $\tau\to K\pi\nu$ decay. In this preliminary result, one loop corrections and soft real photons have been taken into account. As it can be seen in Eqs. (\ref{eq:oneloop},\ref{eq:sp}) the result has a cinematic enhacement ($u$-dependence) and is free of UV singularities.
This result can be used  to improve the estimation of the soft photon corrrection done by numerical simulation, provided that $\omega_{0}$ be small enough compared with typical masses involved in the process. In order to have a more reliable estimation of the MI electromagnetic effects, an inclusive real photon correction must be computed.  In a forthcoming work, the total MI real photon correction (see Ec. (\ref{eq:spb})) will be estimate.
\section*{Acknowledgements}
\nin
This work has been supported in part by the EU MRTNCT-2006-035482 (FLAVIAnet), by MEC (Spain) under Grant FPA2007-60323 and by the Spanish Consolider Ingenio 2010 Programme CPAN (CSD2007-00042). F.V. Flores Ba\'ez  thank AECID (MAEC, Spain) for grant M\'ex:/0288/09  and also thank the LPTA-Montpellier for hospitality.





\begin{thebibliography}{999}
\vspace*{-0.25cm}
\bibitem{sm} A. Pich, QCD Test from Tau-Decay Data, Proceedings of Tau-Charm Factory Workshop (SLAC, Stanford, California,23-27 Mey 1989) SLAC Report-343, (1989), 416.\\
M. Davier, A. H\"{o}cker, Z. Zhang, Rev. Mod. Phys. 78 (2006) 1043.

\bibitem{qcd} E. Braaten, Phys. Rev. Lett. 60, (1988) 1606.\\
S. Narison, A. Pich, Phys. Lett. B 211, (1988) 183.\\
E. Braaten. Phys. Rev. D. 39, (1989) 1458.\\
E. Braaten, S. Narison, A. Pich. Nucl. Phys. B 373, (1992) 581.


\bibitem{aleph} R. Barate \textit{et al}, Eur. Phys. J. C. 11, (1999) 599.

\bibitem{opal} G. Abbiendi \textit{et al}, Eur. Phys. J. C.35, (2004) 437.

\bibitem{babar} BaBar Collaboration.\\ Physical Review D 76, (2007) 051104(R).

\bibitem{belle} Belle Collaboration.\\ Physics Letters B 654, (2007) 65.

\bibitem{antonelli} M. Antonelli \textit{et al}  for the FlaviaNet Working Group on Kaon Decays. Submitted to Eur. Phys. J. C. arXiv:1005.2323[hep-ph]

\bibitem{mirkes} M. Finkemeier, E. Mirkes, Z.Physics C 72, (1996) 619.

\bibitem{jamin} Matthias Jamin, Antonio Pich, Jorge Portoles. Physics Letters B 664, (2008) 78-83.\\
Diogo R. Boito, Rafael Escribano and Matthias Jamin. Eur. Phys. J. C 59, (2009) 821-829.

\bibitem{rchpt} G. Ecker, J. Gasser, A. Pich, E. de Rafael. Nucl. Phys. B 321, (1989) 311.\\
G. Ecker, J. Gasser,H. Leutwyler, A. Pich, E. de Rafael. Phys. Lett. B 223, (1989) 425.

\bibitem{erler} J. Erler. Rev. Mex. Fis. 50, (2004) 200.

\bibitem{ginsberg} Edward S. Ginsberg. Physical Review Vol.142, No. 4, (1966) 1035.\\
Edward S. Ginsberg. Physical Review 162, No. 5, (1967) 1570.

\bibitem{chpt} See a review in A. Pich. Rept. Prog. Phys.58:563-610,1995.

\bibitem{chptconfotones} Res Urech, Nuclear Physics B 433, (1995) 234-254.

\bibitem{chptleptons} M. Knecht, H. Neufeld, H. Rupertsberger, P. Talavera. Eur. Phys. J.C. 12, (2000) 469-478.


\bibitem{kl3} V. Cirigliano, M. Knecht, H. Neufeld and P.Talavera. Eur. Phys. J. C 23, (2002) 121.



\bibitem{photos} E. Barberio, B. van Eijk and Z. Was. Comput. Phys. Commun. 66, (1991) 115.

\bibitem{klor} T.C. Andre. Nucl. Phys. Proc. Suppl. 142, (2005) 58-61. UMI-31-49380.

\bibitem{kl3b} Vincenzo Cirigliano, Maurizio Giannotti, Helmut Neufeld. JHEP 0811:006 (2008).

\bibitem{Yena} D.R. Yennie, S.C. Frautschi and H. Sura. Annals of Physics 13:379-452 (1961).\\
N.Meisteir and D.R. Yennie. Physical Review 130, No. 3, (1963) 1210.

\bibitem{cirigliano} V. Cirigliano, G. Ecker, H. Neufeld. Phys. Lett. B 513, (2001) 361-370.

\bibitem{denner} A. Denner and W. Beenakker. Nucl. Phys. B 338, (1990) 349-370.

\bibitem{bloch} F. Bloch and A. Nordsieck, Phys. Rev. 52, (1937) 54.
\bibitem{low} F.E. Low, Phys. Rev. 110, (1958) 974.
\bibitem{kroll} T.H. Burnett and N.M. Kroll, Phys.Rev.Lett.20, (1968) 86.
\bibitem{coester} F. Coester, Phys.Rev. 83, (1951) 798.\\Jauch and Rohrlich. \textit{The Theory of photons and electrons: the relativistic quantum field theory of charged particels with spin one-half}.\\ Addison-Wesley, 1955.\\
see also for example D.M.Tun, S.R. Juarez and A. Garc\'ia, Phys. Rev. D 40, (1989) 2967.
\bibitem{passemar} Private communication with E. Passemar.






\end{thebibliography}








\end{document}